\documentclass{article}

\usepackage{graphicx}
\usepackage{amsmath}
\usepackage{amssymb}
\usepackage{placeins}
\usepackage{subfigure}
\usepackage{subfiles}
\usepackage{tabularx}
\usepackage{ xcolor,colortbl }
\newcommand{\pink}{\textcolor{black}}
\newcommand{\new}{\textcolor{black}}
\newcommand{\corr}{\textcolor{black}}

\usepackage{url}

\definecolor{Gray}{gray}{0.85}

\newcolumntype{a}{>{\columncolor{Gray}}c}
\newcolumntype{b}{>{\columncolor{white}}c}

\def\href#1{\relax}

\title{Extending MIEZE spectroscopy towards thermal wavelengths}
\author{Johanna K. Jochum$^{1,2}$, Christian Franz$^3$, Thomas Keller$^4$, Christian Pfleiderer$^{2,5}$}
\date{%
    $^1${Heinz Maier\,-\,Leibnitz Zentrum (MLZ), Technische Universit\"at M\"unchen, D-85748 Garching, Germany}\\%
    $^2$Physik Department, Technische Universit\"at M\"unchen, D-85748 Garching, Germany\\
    $^3$ J\"ulich Centre for Neutron Science JCNS\,-\,MLZ, Forschungszentrum J\"ulich GmbH Outstation at MLZ FRM\,-\,II, D-85747 Garching, Germany \\
    $^4$ Max-Planck-Institut f\"ur Festk\"orperforschung, Heisenbergstrasse 1, D-70569, Stuttgart, Germany\\
    $^5$ Centre for QuantumEngineering (ZQE), Technische Universit\"at M\"unchen, D-85748 Garching, Germany\\[5ex]%
    \today
}

\begin{document}

\maketitle                       

\begin{abstract}
We propose a Modulation of intensity with zero effort (MIEZE) set-up for high-resolution neutron spectroscopy at momentum transfers up to 3\,\AA$^{-1}$, energy transfers up to\,$~$20\,meV, and an energy resolution in the $\mu$eV-range using both thermal and cold neutrons. 
MIEZE has two prominent advantages compared to classical neutron spin-echo. 
The first one is the possibility to investigate spin-depolarizing samples or samples in strong magnetic fields without loss of signal amplitude and intensity. 
This allows for the study of spin fluctuations in ferromagnets, and facilitates the study of samples with strong spin-incoherent scattering. 
The second advantage is that multi-analyzer setups can be implemented with comparatively small effort.
The use of thermal neutrons increases the range of validity of \pink{the spin-echo approximation} towards shorter spin-echo times.
In turn, the thermal MIEZE option for greater ranges (TIGER) closes the gap between classical neutron spin-echo spectroscopy and conventional high-resolution neutron spectroscopy techniques such as triple-axis, time-of-flight, and back-scattering. 
To illustrate the feasibility of TIGER we present the details of an implementation at the beamline RESEDA at FRM II by means of an additional velocity selector, polarizer and analyzer. 
\end{abstract}

\section{Introduction}

The need for high-resolution neutron spectroscopy over a wide range of momentum transfers, $Q$, has motivated major advances in neutron instrumentation in recent years.
Key scientific questions that require the high energy resolution over a large $Q$ range concern, e.g., the investigation of ground states without long-range order or well-defined excitations. 
In magnetic materials this includes spin frozen states in classical spin glasses and highly correlated spin states in geometrically frustrated systems \cite{2019_Musgraves_}. 
For instance, such high-resolution spectroscopy over a large dynamic range promises unambiguous identification of quantum spin liquids \cite{paddison_continuous_2017, shen_evidence_2016}, where knowledge of spectral details such as the existence of tiny gaps and the structure of the low-lying modes is key.
Another major field of research where such instrumentation is required concerns hydrogen~--~ or lithium~--~based functionalities in solids. 
Since many batteries rely on ion exchange and/or the ionic conductivity of lithium and hydrogen, determination of diffusion mechanisms in these materials is essential in understanding and improving this performance \cite{hester_neutron_2016, kuznetsov_neutron_2021, li_wassermobilitat_2021, klein_neutron_2021, okuchi_quasielastic_2018}. 
Similarly, the diffusion in ionic liquids and solvent-based electrolytes has been attracting great interest \cite{lundin_structure_2021, burankova_linking_2018, adya_dynamics_2007, osti_microscopic_2019}.

The importance of these questions has been addressed in terms of the development of multianalyzer spectrometers at various beam-lines worldwide.
For instance, CAMEA a modern spectrometer of this type has recently started operation at the Paul Scherrer Institute \cite{Groitl2016, Janas2021, Allenspach2021}. 
In comparison to conventional inelastic neutron scattering (INS) techniques such as time-of-flight or triple axis spectroscopy, neuron spin-echo (NSE) spectroscopy offers several key advantages.

First, in classical INS the \corr{energy} resolution is strongly coupled to the incoming neutron energy and the width of the wavelength band $\frac{\Delta \lambda}{\lambda}$. Increasing the resolution can be achieved by decreasing the incoming neutron energy \corr{and\textbackslash or $\frac{\Delta \lambda}{\lambda}$. This leads to a reduction of Q space coverage and neutron flux.} 
\corr{In NSE, the resolution is not coupled to $\frac{\Delta \lambda}{\lambda}$ since the energy information is encoded in the Larmor precession of the neutron spin \cite{1972Mezei}. However, the energy resolution of NSE does depend on the wavelength used for the measurement (see equation \ref{eq:tau} below). So,} while increasing the incoming neutron energy does reduce the energy resolution, thermal NSE would still offer an energy resolution in the $\mu$eV range for large energy and momentum transfers.
Second, in NSE it is easier to separate diffuse inelastic signal contributions in disordered states from elastic signal contamination by nuclear Bragg scattering.
Third, the direct measurement of the intermediate scattering function \corr{$I(Q, \tau)$} as compared to the scattering function $S(Q, \omega)$ recorded with conventional techniques permits to disentangle the simultaneous presence of processes over a wide dynamic range and provides information on dynamics on the atomic and molecular scales directly amenable to theoretical or numerical modeling, such as molecular dynamics simulations. 
Fourth, in spin-echo methods, the resolution function enters as a product rather than a convolution, thus facilitating resolution correction of data by simply normalizing the raw \corr{data} with the resolution function.
\corr{In general, spin-echo spectroscopy is performed within, but not limited to the spin-echo approximation, which assumes that the energy transfers under investigation are much smaller than the energy of the incoming neutron (roughly $\Delta E \leq 0.1 E_{in}$) \cite{Franz_2019PNCMI}. 
In this conventional case the spin-echo phase is proportional to the energy transfer.
The TIGER spectrometer will be operated in contrast to NSE-TAS without crystal analyzers. This will allow for a large spread of energy transfer, where the simple, linear relation between energy transfer and spin-echo phase is no longer valid. 
In addition large energy transfers lead to a large difference between the wavelength of the incident and scattered neutrons, so that for a fixed scattering angle the momentum transfer is not uniquely defined, but varies with the energy transfer. 
In this case the scattering can be analyzed with the help of numerical simulation of the spectrometer, such as in \cite{2018Haslbeck, Tseng2016}.
} 

An inherent disadvantage of conventional neutron spin-echo spectroscopy arises from its sensitivity to depolarizing effects of the neutron beam.
However, placing all spin manipulation upstream of the sample position in the so-called MIEZE implementation of neutron spin-echo spectroscopy allows to avoid these problems. 
The insensitivity to spin flips facilitates the study of samples containing hydrogen, which shows strong spin-incoherent scattering. 
The study of spin excitations is a successful domain of NSE-TAS \cite{keller2022}. While the study of ferromagnets is in principle possible with Mezei's 'ferromagnetic NSE configuration', it requires additional polarizers up- and downstream the sample leading to a strong loss of the signal intensity. 
Especially for ferromagnets, MIEZE has proven to be a powerful alternative \cite{2018Haslbeck, janoschek_fluctuation-induced_2013}.
However, present-day MIEZE implementations are limited to comparatively small energy and momentum transfers in the range of 0.0005--1.8\,\AA$^{-1}$. 
The potential of MIEZE was demonstrated in recent years in several studies.
\new{In the canonical spin ice system Ho$_2$Ti$_2$O$_7$ we observed excitations that are dominated by a transition within the ground state doublet that is well described within an Orbach model below 50\,K \cite{2017Ruminy, Wendl:Master} and inelastic transitions between excited crystal field levels above 50\,K \cite{2019Franz2, ehlers_evidence_2004}.
The dynamical properties in the reentrant spin-glass system Fe$_x$Cr$_{1-x}$ could be tracked across the quantum phase transition between itinerant electron ferro- and antiferromagnetism \cite{Benka2022, Steffen2022}.
The spin dynamics of Fe and Ni was studied above and below its Curie temperature, where spin waves were resolved below T$_c$. In comparison, above T$_c$ critical slowing down of ferromagnetic fluctuations was observed \cite{2015Kindervater, 2019Saubert}. 
In the superconducting ferromagnet UGe$_2$, a cross-over between dynamical properties characteristic of local-moment and itinerant-electron magnetism could be  identified\cite{2018Haslbeck}.
RESEDA was used in a study of the fluctuation-induced phase transitions in cubic chiral magnets, highlighting the complexities of the phase diagrams of these compounds, such as the weak crystallization of skyrmion textures \cite{2019Kindervater, 2019Martin,2022Weber, 2020Kindervater, 2019Franz2}.
A recent highlight concerned the study of emergent Landau levels in MnSi: 
The motion of a spin excitation across topologically non-trivial magnetic order leads to the emergence of Landau levels in the excitations spectrum for magnons in such systems. 
To demonstrate the existence of these energy levels in MnSi unpolarized neutron time-of-flight spectroscopy was used to get an overview of the spectra while polarized three axis spectroscopy probed the distribution of excitation energies and scattering intensities across a large number of positions in parameter space. MIEZE was then used to resolve and confirm the existence of individual Landau levels including those at the lowest energies \cite{2022Weber}.}

Motivated by the wide range of science cases requiring high-resolution data in the regime of large energy and momentum transfers, we propose the extension of the MIEZE technique towards neutron wavelengths as low as 2\,\AA, thus increasing the accessible parameter space substantially.
The thermal MIEZE for greater ranges (TIGER) we propose, will boost the accessible $Q$- as well as energy range of neutron resonance spin-echo spectroscopy. 
Originally conceived some 30 years ago by Golub, G{\"a}hler and Keller \cite{1987Golub, 1990Keller, 1992Gaehler} major progress has been achieved at the beamline RESEDA at the FRM\,II in recent years. 
In turn, we present the specific requirements to implement TIGER as an upgrade of RESEDA \cite{2019Franz, reseda, Haussler2007}.

The outline of our paper is as follows.
In section \ref{sec:state} the state of the art of the MIEZE technique and its implementations at various neutron sources worldwide is summarized.
The technical requirements of thermal MIEZE are presented in section \ref{sec:general}.
This includes a critical assessment of the implications an operation with thermal neutrons will have on the different instrument components, such as polarizer, analyzer, velocity selector, rf-circuits and detector.
We then go on to exploring RESEDA as a possible candidate for the implementation of TIGER.
Our findings are summarized in section \ref{sec:conclusion}.


\section{State of the Art}\label{sec:state}



\begin{figure}
\centering
\includegraphics[width=0.88\textwidth]{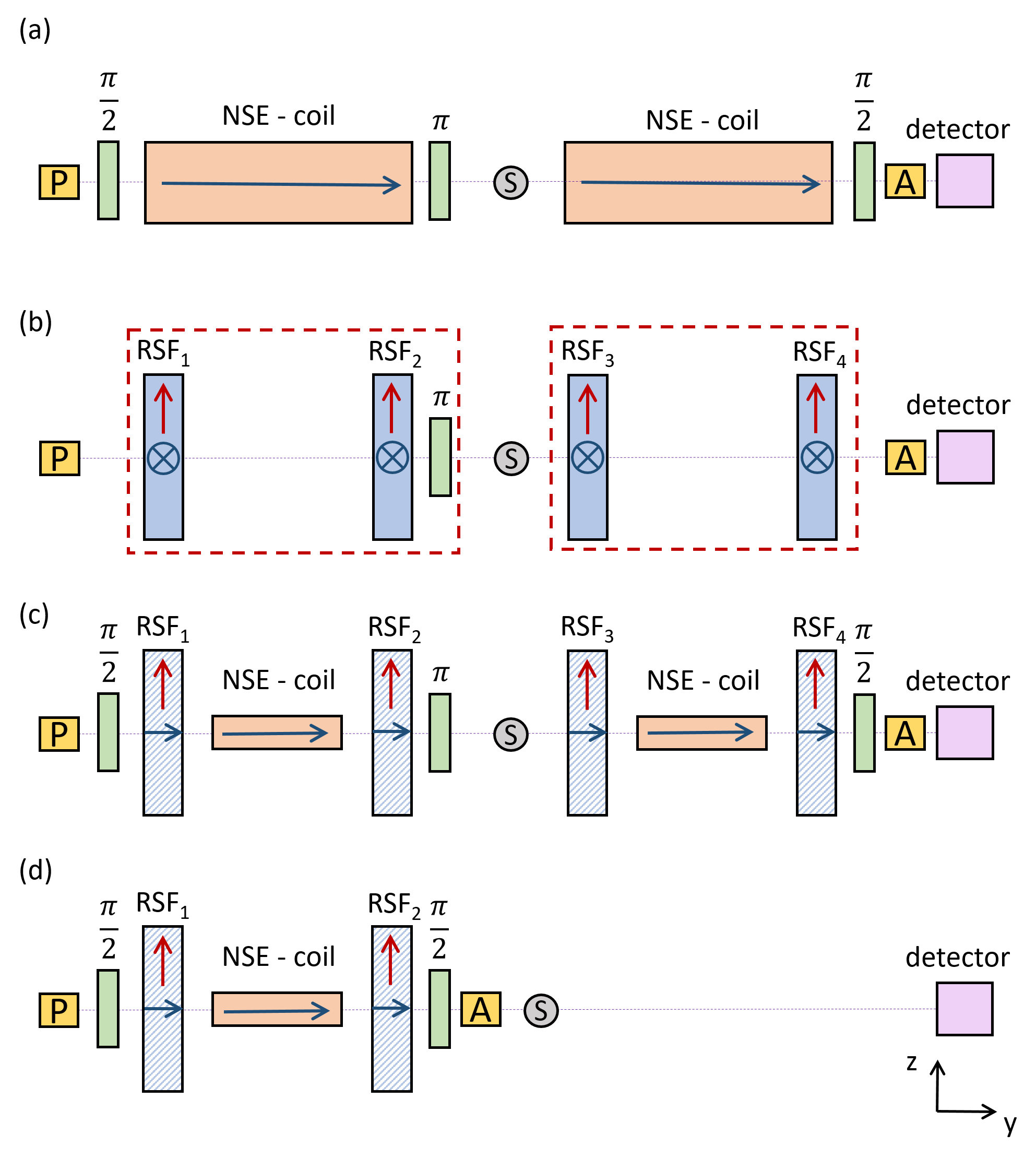}
\caption{Overview of the different neutron spin-echo techniques. P, S and A correspond to polarizer, sample and analyzer respectively. (a) classical neutron spin-echo: the spin precession occurs in large solenoids (NSE-coil); the magnetic fields points along the $y$-direction \cite{1972Mezei}; (b) transverse neutron resonant spin-echo: the solenoids are replaced by pairs of resonant spin-flippers (RSF); here the static field (red) points in the $x$-direction whereas the oscillating field points along $z$; the precession zone is contained in a mu-metal shielding (red dashed line) to avoid depolarization \cite{1987Golub}; (c) longitudinal neutron resonant spin-echo: akin to classical NSE the static fields of the RSF points along $y$, while the oscillating field points along $z$; no mu-metal shielding is needed; a smaller solenoid (NSE-coil) is inserted in between the RSF operating as a field subtraction coil \cite{2011Haeussler}; (d) longitudinal MIEZE: the set-up is analogue to the primary spectrometer arm of the longitudinal NRSE where the analyzer is moved upstream the sample; no more spin manipulation occurs after the sample \cite{reseda}.}
\label{fig:nse}
\end{figure}

The MIEZE technique was developed as a spin-echo variant insensitive to depolarizing conditions at the sample position.
The first implementations of MIEZE followed the design of the (transverse) neutron resonant spin-echo setups where the large solenoids used in classical NSE are replaced by a pair of rf-flippers (see figure \ref{fig:nse}\, (a) and (b)).
\new{In general, a MIEZE setup consists of a wavelength selector (not shown), a polarizer (P), the spin precession zone, which comprises two rf-flippers (RSF$_1$, RSF$_2$) and is followed by the analyzer (A), a sample (S) and a detector. }
In the transverse geometry the fields in the rf-flippers are arranged such, that both the static and oscillating field are perpendicular to the neutron beam.
This has the disadvantage that the neutron beam needs to pass through several millimeters of aluminum to traverse the static field coils.
\new{Additionally, the rf-flipper is enclosed by the static field coil, which leads to a dampening of the resonant circuit for large ($\leq$\,1\,MHz) frequencies \cite{cook_concepts_2014, Jochum2021}. 
To avoid neutron beam depolarization in this geometry, it is necessary to shield the precession zones with mu-metal \cite{Martin2014}.
While in classical NSE the resolution limit originates in the difficulty of achieving highest field homogeniety with increasing field integral,  the resolution of transverse NRSE is limited by path length differences of neutrons traversing the rf-flippers on trajectories that are not parallel to its symmetry axis \cite{2017Martin}. }
\new{These challenges were met by H{\"a}ussler et al. \cite{2005Haeussler, 2003Haeussler} who proposed a longitudinal geometry for NRSE and MIEZE, combining the advantages of classical NSE with NRSE (see figure \ref{fig:nse}\,(c) and (d)).
In this geometry the static field is produced by compact solenoids that do not hinder the neutron beam propagation and point along the neutron beam trajectory, akin to classical NSE.
This offers the additional advantage that the polarization may be maintained by small longitudinal guide fields, and the mu-metal shielding can be omitted. 
The rf-field coils are the same as in the transverse setup \cite{cook_concepts_2014, Jochum2021}. 
The field points perpendicular to the beam, such that the neutron beam passes through the winding of the coil, which consists of aluminium flat wire less than one millimeter thick. 
The longitudinal resonant spin-flipper has the advantage that field inhomogenieties at the field boundaries are symmetric with respect to the rf-flipper and therefore cancel out instead of contributing to the accumulated neutron phase \cite{2016Krautloher}. 
To avoid stray fields from the static fields, the longitudinal geometry offers the option to install a pair of cut-off coils with field direction opposite to the main dc-coils in a Helmholtz configuration around the resonant flipper.}

\new{The first transverse NRSE set-up for user operation were the MUSES instrument at the Orph\'ee reactor in Saclay and the NRSE option for the TAS FLEX at HZB.
Until the reactor shutdown in 2019 the instrument MUSES offered both an NSE and a transverse NRSE option \cite{longeville_spectroscopie_2000, koppe_performance_1996}.
The first MIEZE experiments were performed at MUSES as well \cite{koppe_performance_1996}.
The MIEZE option did, however, never go into user operation. 
The spectrometer RESEDA at the FRM\,II was initially constructed as a transverse NRSE set-up with a multi detector option \cite{Haussler2007} and was later reconstructed to offer the transverse NRSE option. 
RESEDA was rebuilt in its longitudinal configuration at the end of 2014.}

Three groups worldwide work on the development of transverse MIEZE spectrometers at pulsed neutron sources. 
The MIEZE Setup at BL06 VIN ROSE at J-Parc is in full user operation.
It offers measurements in a Fourier time range from 1 to 500\,ps. \cite{hino_current_2013, funama_study_2021, nakajima_crystallization_2020,oda_tuning_2020, oda_phase_2021}
A MIEZE setup has been developed for the Larmor spectrometer at the ISIS neutron source. 
Benchmarking and first experiments on the dynamics of water have been performed \cite{kuhn_time--flight_2021, geerits_time_2019}.
A MIEZE setup at ORNL was recently commissioned.
Seminal work on entangled neutron beams has been carried out with this set-up \cite{shen_unveiling_2020, kuhn_neutron-state_2021}.
However this setup does not have a dedicated beamline \cite{kuhn_neutron-state_2021, lu_operator_2020, shen_unveiling_2020, Zhao2015, 2020Dadisman}.



\begin{figure}
\centering
\includegraphics[width=1\textwidth]{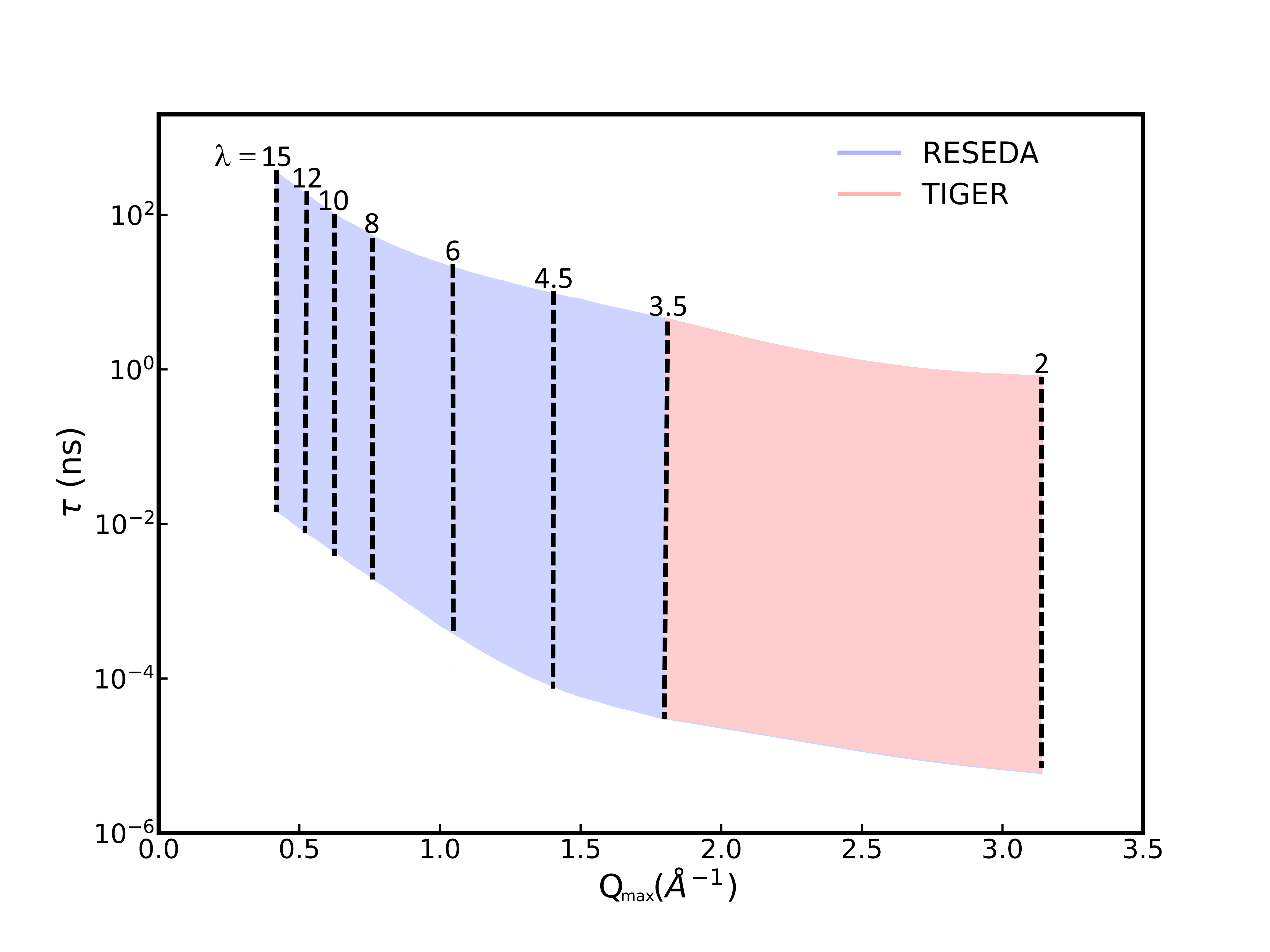}
\caption{Accessible parameter range as a function of $\lambda$, Q$_{max}$ and $\tau$. The blue shaded area marks the parameter space currently available at RESEDA, while the red shaded area shows the additional parameter space that would become available with the TIGER upgrade.}
\label{fig:range}
\end{figure}

RESEDA is the only longitudinal neutron resonance spin-echo spectrometer currently in user operation.
It is situated at the cold neutron guide NL5-S in the Neutron Guide Hall West at the Forschungs-Neutronenquelle Heinz Maier-Leibnitz. 
\corr{In its present configuration the instrument provides access to a range of Fourier times from 0.07\,ps (for $\lambda$\,=\,4.5\,\AA\,) to 11.5\,ns (for $\lambda$\,=\,10\,\AA\,) \cite{2019Franz, 2019Franz3} and a range of scattering vectors from 0.0005 (for $\lambda$\,=\,6\,\AA\,) to 1.4\,\AA$^{-1}$ (for $\lambda$\,=\,4.5\,\AA\, at a maximum scattering angle of $2\theta$ = 60\,$^\circ$).}
\corr{The relationship between Fourier time, wavelength, and energy resolution are as follows: 
\begin{equation}
    \tau = \frac{2m^2}{h^2}\cdot L_{SD}\cdot\Delta f \cdot \lambda^3,
    \label{eq:tau}
\end{equation}
where $m$ is the neutron mass, $h$ is Planck’s constant, $L_{SD}$ is the sample-detector distance, $\Delta f$ is the frequency difference of the two flippers, $\lambda$ is the neutron wavelength and $\Gamma$ is the quasi-elastic energy linewidth.}\\
As part of the reconstruction of RESEDA the radio-frequency circuits were redesigned \cite{Jochum2021}, 
and a field subtraction coil \cite{2003Haeussler, 2019Jochum} was implemented extending the dynamic range towards shorter Fourier times. 
These developments have unlocked the parameter space shaded in blue in figure \ref{fig:range}, which can be set into context of operational spin-echo instruments, a selection of which is summarized in table \ref{tab:tech_data}.

\begin{table}
\hspace*{-1.8cm}
\begin{tabular}{|l|c|c|c|c|c|c|}
\multicolumn{7}{c}{Instrument parameters}\\
\hline
    Instrument  &  $\lambda$ range & $\tau$ range & Q range & Detector & pixel & flux \\
    & & & &size&size&\\
    \hline
    &\AA & ns & \AA$^{-1}$ & cm$^2$ & cm$^2$ & 10$^6$ n cm$^{-1}$s$^{-1}$\\
    \hline
    RESEDA & 3.5--15 & 0.0001--45 & 0.0005--1.8 & 20$\times$20 & 0.156$\times$0.156 & 20 @ 6\AA\\
    TIGER & 2--15 & 0.00003--45 & 0.0005--3.0 & 20$\times$20 & 0.156$\times$0.156 & 20 @ 6\AA\\
    JNSE & 4.5--16 & 0.002--500 & 0.02--1.8 & 32$\times$32 & 1$\times$1 & 10 @ 7.0\AA\\
    IN15 & 6--25 & 0.004--952  & 0.02--1.21 & 32$\times$32 & 1$\times$1 & 20 @ 7\AA\\
    SNS-Spin-Echo & 2--14 & 0.08--280 & 0.03--3.1 & 30$\times$30 & 1$\times$1 & 1 @ 7.0\AA\\
    CHRNS NSE & 4.5--15 & 0.003--200 & 0.02--1.8 & 32$\times$32 & 1$\times$1 & 31 @ 5.0\AA\\
    BL06 Vin Rose & 3--13 & 0.001--0.5 & 0.2--1.0 & 32$\times$32 & 0.1$\times$0.1 & 31 @ 5.0\AA\\
    \hline 
\end{tabular}
\caption{Present key parameters of RESEDA \cite{2019Franz}, and other MIEZE and spin-echo setups worldwide: J-NSE \cite{JNSE}, IN15 \cite{IN15_2}, SNS-Spin-echo \cite{SNSNSE}, CHRNS NSE \cite{CHRNSNSE}, BL06 Vin Rose \cite{BL06VinRose}; The $\tau$ range describes the full dynamic range, taking all available wavelengths into account; The flux is listed for a representative wavelength;}
\label{tab:tech_data}
\end{table}

Currently, three longitudinal MIEZE spectrometers
are under development at reactor sources:
(1) A longitudinal neutron resonance spin-echo spectrometer combined with a MIEZE configuration is planned at the C33 beam port at China Mianyang Research Reactor (CMRR) \cite{liu_resolution_2020};
(2) The SEM spectrometer at the PIK reactor complex is planned as a longitudinal neutron resonant spin-echo spectrometer with a MIEZE option \cite{kovalchuk_instrument_2021};
(3) Furthermore, a SANS setup with MIEZE option is being installed at the ILL \cite{SAM}.\\

\section{Thermal MIEZE for greater ranges (TIGER)}\label{sec:general}



\new{
The following requirements need to be met for the implementation of TIGER:
(i) The availability of a high neutron flux down to $\lambda$\,=\,2\,\AA\,. This includes a source offering the desired wavelengths spectrum and a velocity selector that can reach wavelength down to 2\,\AA\, while still maintaining a $\frac{\Delta \lambda}{\lambda}$ of 0.1-0.2.
While a broader wavelength band would mean a higher neutron flux, the \corr{width of the MIEZE group} which corresponds to the Fourier transform of the wavelength band, would decrease for a larger wavelength band \cite{2019Jochum}. 
This becomes relevant for long Fourier times (high energy resolution), where the \corr{width of the MIEZE group} shrinks to below 1\,cm, which leads to an inefficient operation of the presently used CASCADE detector \cite{2019Jochum}.
(ii) The neutron optics elements, such as neutron guide system, polarizer and analyzer need to work efficiently at wavelengths $\leq$\,2\,\AA\,.
(iii) Since the current required to perform a $\pi$-flip is indirectly proportional to the neutron wavelength, the rf-flippers will need to support higher currents \cite{cook_concepts_2014, Jochum2021}.  
This is especially important for frequencies above 1\,MHz where an increase in Ohmic resistance due to the skin effect makes it increasingly difficult to feed power into the rf-flipper.
(iv) It is necessary that the neutron detector still offers sufficient sensitivity to neutrons with shorter wavelengths.
}



\begin{figure}
\centering
\includegraphics[width=1\textwidth]{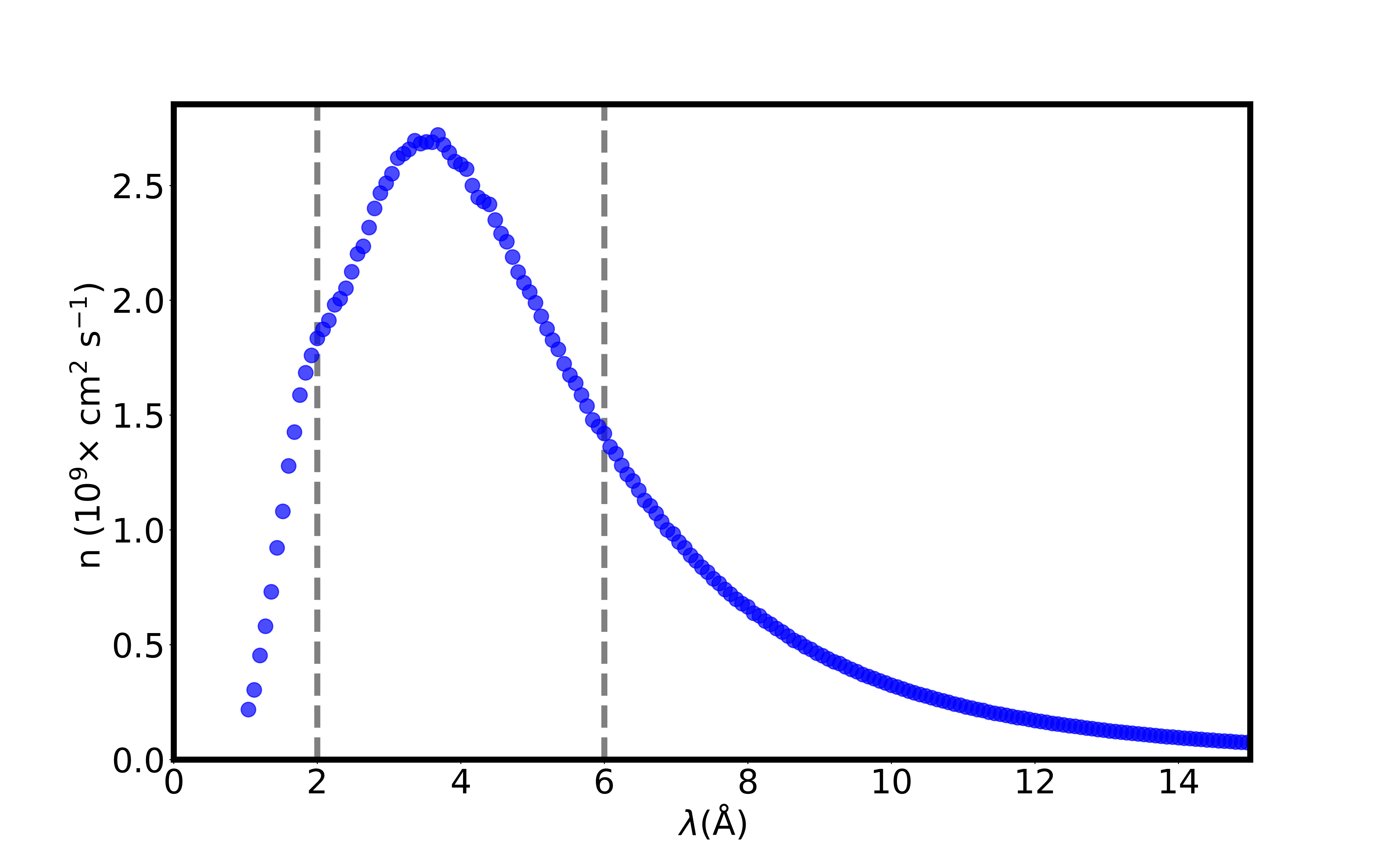}
\caption{Neutron flux at the neutron guide NL5-S, at the position of RESEDA prior to the velocity selector.}
\label{fig:source}
\end{figure}

While RESEDA has so far been optimized for the operation with cold neutrons, which do facilitate measurements with higher energy resolution and small Q, the spectrum available at NL5-S offers neutrons with wavelengths down to 1.5\AA, making operations down to 2\,\AA\, feasible (see figure \ref{fig:source}).
Compared to a wavelength of 6\,\AA, a standard wavelength used at RESEDA offering a good compromise between resolution and flux, the flux at 2\,\AA\, is 15\% higher (see figure \ref{fig:source}).

\begin{figure}
\centering
\includegraphics[width=1\textwidth]{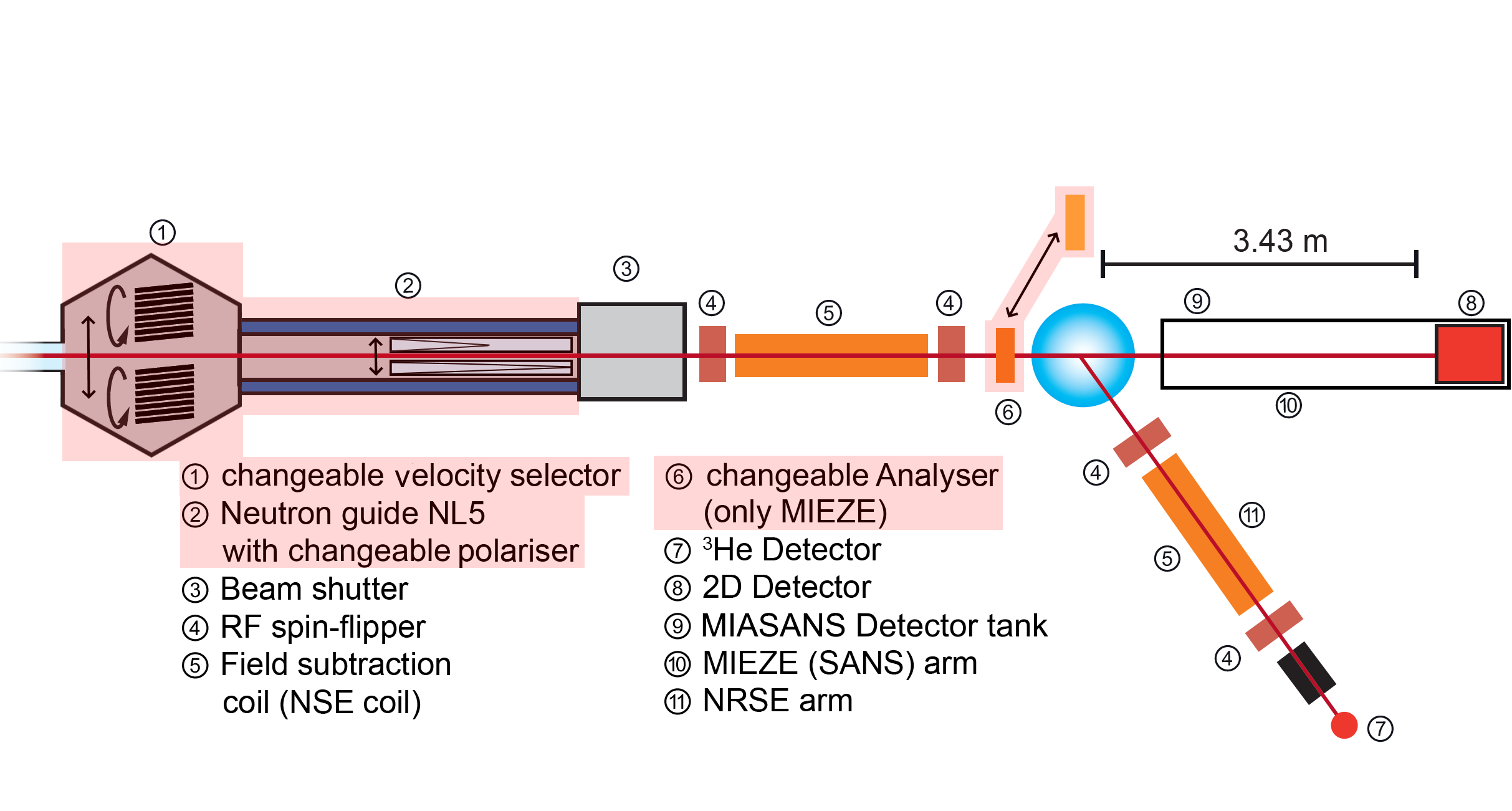}
\caption{Schematic of the proposed setup for thermal neutrons, TIGER at the beamline RESEDA. Components of RESEDA to be upgraded in TIGER are highlighted in light red.}
\label{schema}
\end{figure}

The implementation of TIGER at RESEDA will require the installation of an additional polarizer, analyzer and velocity selector as well as a translation systems for automated switching between the velocity selectors and polarizers for thermal and cold neutrons to guarantee the availability of both options for the user community.
The Q-$\tau$ range that would be added to the current dynamic range of RESEDA is shaded in red in Fig.\,\ref{fig:range}.
Figure \ref{schema} shows a schematic of RESEDA, where the elements that will require modification for the proposed upgrade are shaded in red.

\begin{figure}
\centering
\includegraphics[width=1\textwidth]{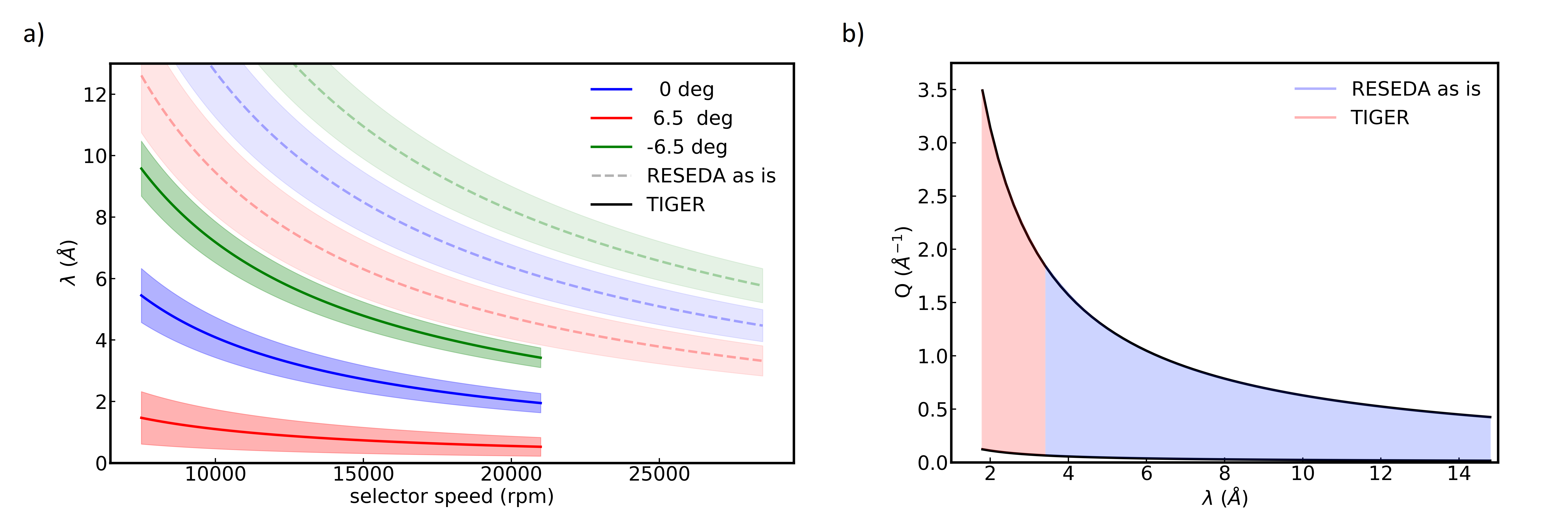}
\caption{Parameter range of the velocity selector of TIGER as compared with the existing selector. (a) Newly accessible wavelength range as a function of selector speed in rotations per minute for different tilt angles of the selector. (b) Newly accessible Q range as a function of wavelength.}
\label{fig:selector}
\end{figure}

A thermal velocity selector will ensure operation of RESEDA down to $\lambda$ = 2\,\AA. 
It is possible to construct a wavelength selector with the desired $\lambda$ and $\frac{\Delta\lambda}{\lambda}$ at a slightly reduced transmission (91\% as compared to 98\% for the current velocity selector). 
\corr{The specifications of this selector include a screw angel $\alpha$\,=\,15.5$^\circ$, a blade height of 60\,mm, 144 blades, and a maximum rotation speed of 21000\,rpm.}
Shown in Fig. \ref{fig:selector} a) is a comparison between the existing wavelength selector and the additional selector that would be required for TIGER. 
The central lines denote the central wavelength, whereas the shaded areas depict the spread of the wavelength band. 
The figure further shows the wavelength range and $\frac{\Delta\lambda}{\lambda}$ for different tilt angles of the selector. 
Both strongly depend on the tilt angle which allows for a greater versatility of a single selector.

Similarly to the current polarizer at RESEDA, a polarizing cavity consisting of two parallel channels, each with two V cavities in series, and a total length of 2040\,mm \cite{Swiss} would be a suitable polarizing device. 
The polarizer would work efficiently in the wavelength range from 1.8\,to\,7.9\,\AA\,\cite{Schanzer_2016}.
The channels comprise 0.3\,mm Si substrates with an m\,=\,5 Fe/Si coating.
The outer dimensions of the cavity and the magnetic field strength (45-50\,mT) would be the same as for the one currently installed at RESEDA \cite{2019Franz}, facilitating the switching between both cavities.

As a neutron polarization analyzer device a transmission bender with integrated 80\,' collimator would be best suited \cite{analyzer}. 
Using Fe/Si m\,=\,5 supermirrors reduces the length of the bender to 68\,mm, twice the size of the current device used for cold neutrons. 
The lammelar collimator that follows the analyzer to absorb the reflected beam, would be 12.9\,mm long and consist of Si wafers with a Gd coating. 
Since the bender is mounted such that it can be exchanged manually at any time this increase in size does not cause any inconveniences. 

\begin{figure}
\centering
\includegraphics[width=1.0\textwidth]{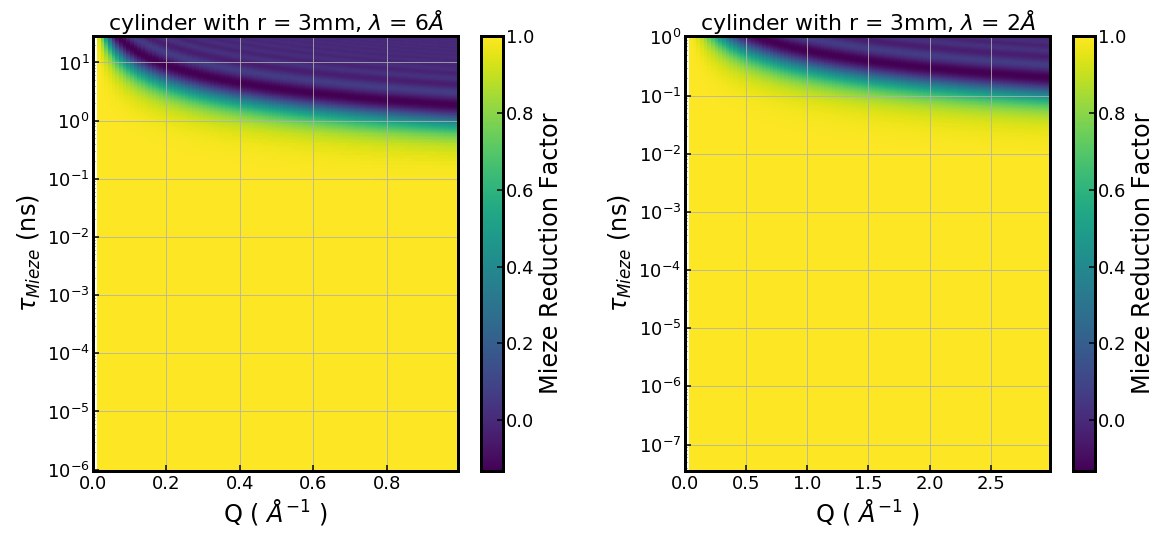}
\caption{Reduction factor for a cylindrical sample for RESEDA and TIGER. The yellow shaded areas denotes a perfect signal. In the blue-shaded areas the MIEZE signal is lost due to path length differences of the neutrons.}
\label{fig:redction}
\end{figure}

A major advantage of the longitudinal NRSE and MIEZE set-up concerns the self-correction of path length differences of parallel neutron trajectories, and a strongly reduced sensitivity to differences of divergent beam trajectories \cite{2016Krautloher}. 
The signal reduction depends mostly on the geometry of the sample under investigation. 
RESEDA offers an online reduction factor calculator \cite{RedCalc} so users can find the optimal sample shape. 
Shown in Fig. \ref{fig:redction} (b) is the calculated signal reduction as a function of wavelength for path length differences for a typical sample (single crystal cylinder with a radius of 3\,mm), up to the maximum scattering angle 2\,$\theta$ of 60\,$^{\circ}$. 
In the yellow shaded area the signal contrast is not affected by the sample size. 
In the blue-shaded areas the MIEZE contrast is reduced due to path length differences of the neutrons. 
For this sample and a wavelength of 6\,\AA, at a maximum Q of 1\,\AA$^{-1}$ a Fourier time of 1\,ns can be reached, which corresponds to an energy resolution of 0.66\,$\mu$eV.
In contrast, for the same sample, a wavelength of 2\,\AA, and a maximum Q of 3\,\AA$^{-1}$ a Fourier time of 0.1\,ns can be reached, which corresponds to an energy resolution of 6.58\,$\mu$eV.
The figure further shows that the resolution is higher in the small angle regime, where the full dynamic range of RESEDA can be accessed with almost any sample shape. 

While the additional polarizer, analyzer and selector are the key components for this instrument upgrade, we would like to discuss the performance of the current detector and resonating circuits at wavelengths down to 2\,\AA. 
RESEDA currently uses a CASCADE detector \cite{2016Koehli} with 6 $^{10}B$ detection foils.
The detection efficiency for neutrons with a wavelength of 6\,\AA\, is 63\,\%. 
For $\lambda$\,=\,2\,\AA\, neutrons this will be reduced to 34\,\%.
The introduction of the new superconducting static field coils requires a better detector than the CASCADE type currently used at RESEDA \cite{2016Koehli, Jochum2021, Jochum2022}.
Therefore, we have started exploring different detector technologies for the application in MIEZE.
One possible candidate would be a scintillation detector based on the Timepix chip, which is currently being tested for neutrons by Losko et al. \cite{losko_new_2021}.

The rf-coils used at RESEDA have been optimized for cold neutrons and high energy resolution.
Once the superconducting dc-coils have been commissioned with neutrons the highest available Fourier time at 6\,\AA\, will be 23.1\,ns (107\,ns at 10\,\AA). 
To achieve a resonant spin-flip at shorter wavelengths than currently at use at RESEDA, it is necessary to operate the rf-flippers at higher currents.
The current flipper design will need to be revised to perform spin-flips of $\lambda$\,=\,2\,\AA\, neutrons at the highest available flipper frequencies.
Reducing the area and increasing the thickness of the rf-flipper will reduce the current needed for the spin-flips and should make it possible to achieve an energy resolution of 7\,$\mu eV$ for $\lambda$\,=\,2\,\AA\, neutrons\cite{cook_concepts_2014}. 

As RESEDA offers several modes of operation (NRSE, MIEZE, MIEZE-SANS) the integration of TIGER should be such that an easy change between the different instrumental configurations is guaranteed, including a system that permits switching automatically between the different selectors and different polarizing cavities. 

\section{Conclusion}\label{sec:conclusion}

We propose the implementation of a thermal MIEZE setup for high-resolutions spectroscopy at high energy and momentum transfers, at the neutron resonant spin-echo spectrometer RESEDA at the Heinz Maier-Leibnitz Center in Garching, Germany.
The proposed developments are driven by a the need of the scientific community for high resolution measurements over a wide range in Q, which is needed for the investigation of new exotic ground states without long range order. 
The implementation will allow for an energy resolution of 6.58\,$\mu$eV at Q values of 3\,\AA$^{-1}$ and energy transfers up to 20.5\,meV. 

\section{Acknowledgements}
We thank Jon Leiner and Peter B\"oni for fruitful discussions.
This research was funded through the BMBF projects ‘Longitudinale Resonante Neutronen Spin-Echo Spektroskopie mit Extremer Energie-Aufl\"osung’ (F\"orderkennzeichen 05K16W06) and ‘Resonante Longitudinale MIASANS Spin-Echo Spektroskopie an RESEDA’ (F\"orderkennzeichen 05K19W05)

\newpage

\bibliographystyle{plain}
\bibliography{bib}

\end{document}